\documentclass[twocolumn,superscriptaddress]{revtex4}
\usepackage{graphicx}
\usepackage{dcolumn}
\usepackage{amsmath}
\usepackage{amsfonts}
\usepackage{amssymb}
\begin{document}

\title{\textbf{The power of patience: A behavioral regularity in limit order placement}}

\author{Ilija Zovko}
\email{zovko@santafe.edu}
\affiliation{Santa Fe Institute, 1399 Hyde Park Rd., Santa Fe NM 87501}
\affiliation{CeNDEF, University of Amsterdam, Roetersstraat 11, Amsterdam, The Netherlands}


\author{J. Doyne Farmer\footnote{McKinsey Professor}}
\affiliation{Santa Fe Institute, 1399 Hyde Park Rd., Santa Fe NM 87501}
\email{jdf@santafe.edu}

\date{May 1, 2002}

\begin{abstract}
In this paper we demonstrate a striking regularity in the way
people place limit orders in financial markets, using a data set
consisting of roughly seven million orders from the London Stock
Exchange.  We define the relative limit price as the difference
between the limit price and the best price available.  Merging the
data from 50 stocks, we demonstrate that for both buy and sell
orders, the unconditional cumulative distribution of relative
limit prices decays roughly as a power law with exponent
approximately $-1.5$.  This behavior spans more than two decades,
ranging from a few ticks to about 2000 ticks. Time series of
relative limit prices show interesting temporal structure,
characterized by an autocorrelation function that asymptotically
decays as $C(\tau) \sim \tau^{-0.4}$.  Furthermore, relative limit
price levels are positively correlated with and are led by price
volatility.  This feedback may potentially contribute to clustered
volatility.

\end{abstract}
\maketitle

In this paper we demonstrate a new behavioral regularity of
financial markets. Most modern financial markets use a continuous
double auction mechanism, based on limit orders, which specify
both a quantity and a limit price (the worst acceptable price). We
study the {\it relative limit price} $\delta (t)$, the limit price
in relation to the current best price.  For buy orders $\delta (t)
= b(t) - p(t)$, where $p$ is the limit price, $b$ is the best bid
(highest buy limit price), and $t$ is the time when the order is
placed.  For sell orders $\delta (t) = p(t) - a(t)$, where $a$ is
the best ask (lowest sell limit price)\cite{foot1}.

The limit order trading mechanism works as follows: As each new limit
order arrives, it is matched against the queue of pre-existing limit
orders, called the {\it limit order book}, to determine whether or not
it results in any immediate transactions.  At any given time there is
a best buy price $b(t)$, and a best ask price $a(t)$.  A sell order
that crosses $b(t)$, or a buy order that crosses $a(t)$, results in at
least one transaction.  The matching for transactions is performed based on
price and order of arrival.  Thus matching begins with the order of
the opposite sign that has the best price and arrived first, then
proceeds to the order (if any) with the same price that arrived
second, and so on, repeating for the next best price, etc.  The
matching process continues until the arriving order has either been
entirely transacted, or until there are no orders of the opposite sign
with prices that satisfy the arriving order's limit price.  Anything
that is left over is stored in the limit order book.

Choosing a relative limit price is a strategic decision that involves
a tradeoff between patience and profit. Consider, for example, a sell
order; the story for buy orders is the same, interchanging ``high''
and ``low''.  An impatient seller will submit a limit order with a
limit price well below $b(t)$, which will typically immediately result
in a transaction.  A seller of intermediate patience will submit an
order with $p(t)$ a little greater than $b(t)$; this will not result
in an immediate transaction, but will have high priority as new buy
orders arrive.  A very patient seller will submit an order with $p(t)$
much greater than $b(t)$.  This order is unlikely to be executed
soon, but it will trade at a good price if it does.  A
higher price is clearly desirable, but it comes at the cost of
lowering the probability of trading -- the higher the price, the lower
the probability there will be a trade.  The choice of limit price is a
complex decision that depends on the goals of each agent.  There are
many factors that could affect the choice of limit price, such as the
time horizon of the trading strategy.  {\it A priori} it is not
obvious that the unconditional distribution of limit prices should
have any particular simple functional form.

We investigate the relative limit price $\delta(t)$ for stocks
traded on the London Stock Exchange between August 1, 1998 and
April 31, 2000.  This data set contains many errors; we chose the
names we analyse here from the several hundred that are traded on
the exchange based on the ease of cleaning the data, trying to
keep a reasonable balance between high and low volume stocks
\cite{foot2}. This left 50 different names, with a total of
roughly seven million limit orders.  We discard orders with
negative values of $\delta$, i.e., we consider only orders that
are placed outside the spread \cite{foot3}.
\begin{figure}[ptb]
   \begin{center}
   \includegraphics[scale=0.37, angle=-90]{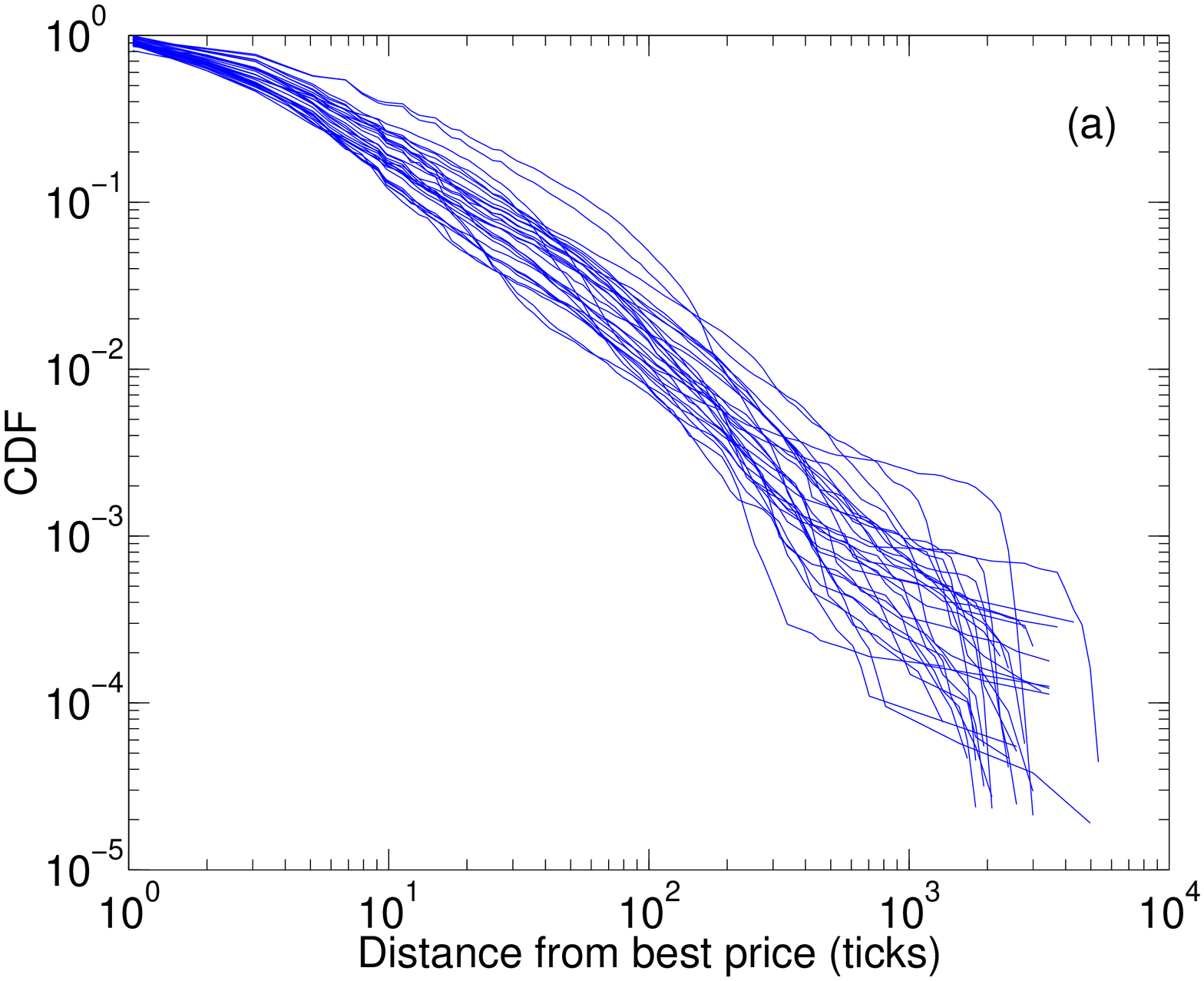}
   \includegraphics[scale=0.37, angle=-90]{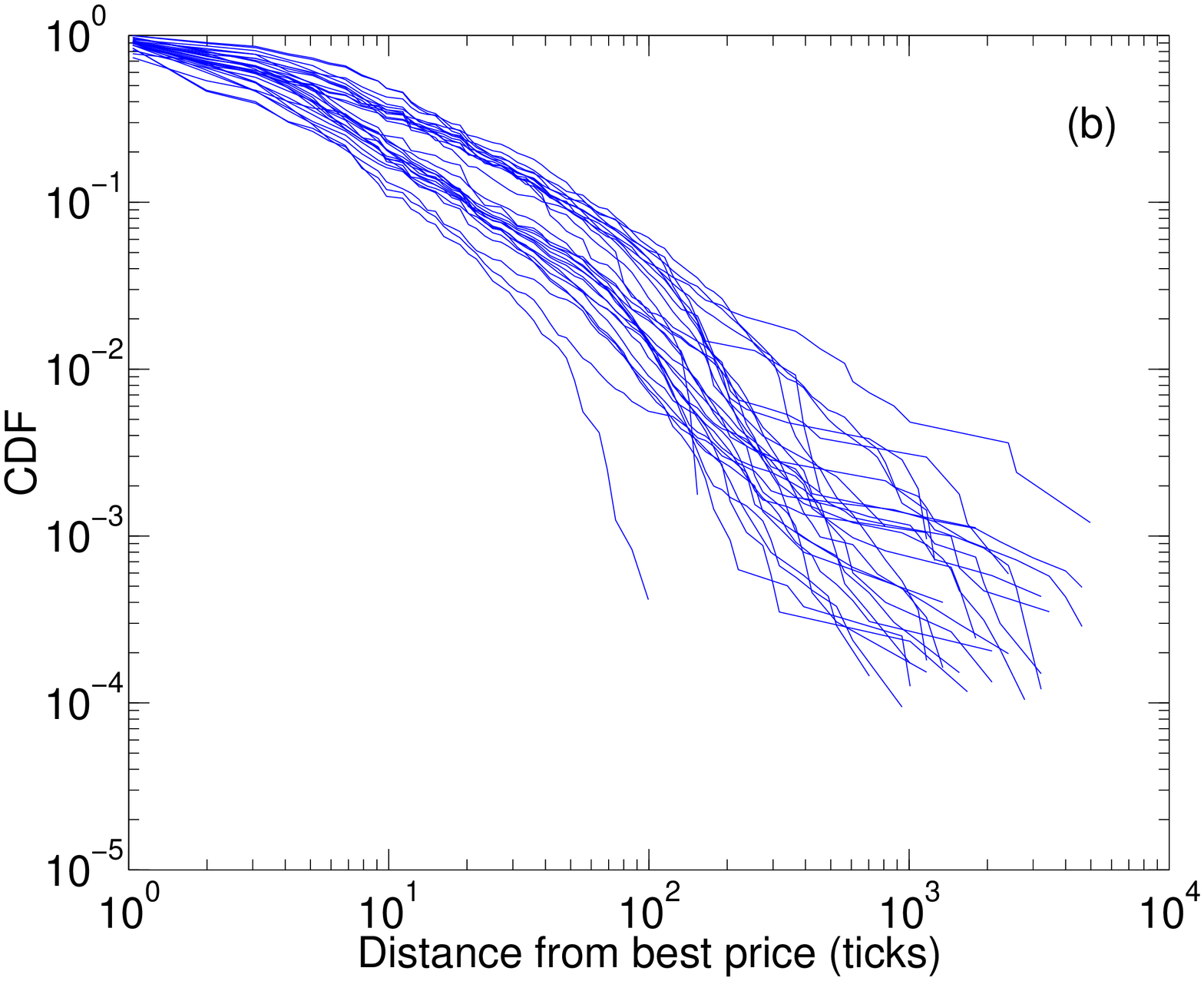}
   \caption{
   (a) Cumulative distribution functions
   $P(\delta) = \mbox{Prob} \{ x \geq \delta \}$ of
   relative limit price $\delta$ for both buy and sell orders for the
   15 stocks with the largest number of limit orders during the period
   of the sample (those that have between 150,000 and 400,000 orders
   in the sample.)  (b) Same for 15 stocks with the lowest number of
   limit orders, in the range 2,000 to 100,000. \cite{foot4} }
   \label{uncondHV}
    \end{center}
\end{figure}
Figure (\ref{uncondHV}) shows examples of the cumulative
distribution for stocks with the largest and smallest number of
limit orders.  Each order is given the same weighting, regardless
of the number of shares, and the distribution for each stock is
normalized so that it sums to one.  There is considerable
variation in the sample distribution from stock to stock, but
these plots nonetheless suggest that power law behavior for large
$\delta$ is a reasonable hypothesis.  This is somewhat clearer for
the stocks with high order arrival rates.  The low volume stocks
show larger fluctuations, presumably because of their smaller
sample sizes. Although there is a large number of events in each
of these distributions, as we will show later, the samples are
highly correlated, so that the effective number of independent
samples is not nearly as large as it seems.

\begin{figure}[ptb]
    \begin{center}
    \includegraphics[scale=0.37, angle=-90]{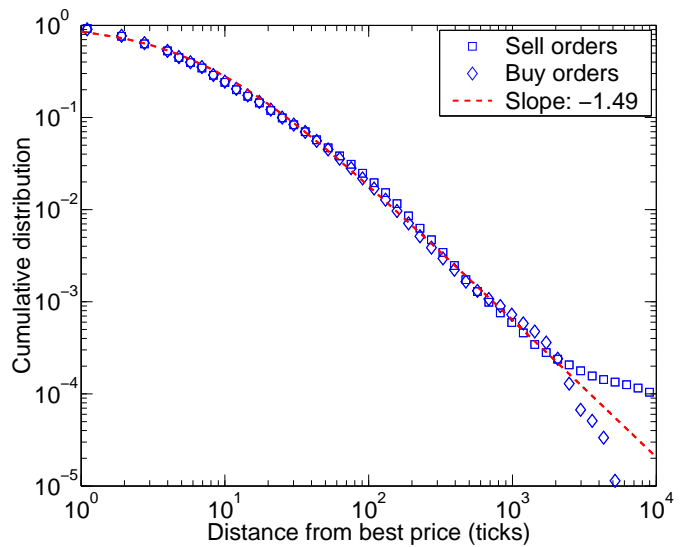}
    \caption{ An estimate of the cumulative probability distribution
        based on a merged data set, containing the relative limit order
        sizes $\delta(t)$ for all 50 stocks across the entire sample.  The
        solid curve is a non-linear least squares fit to the logarithmic
        form of equation (\ref{probDist}).}
    \label{powerlaw}
    \end{center}
    \end{figure}

To reduce the sampling errors we merge the data for all stocks, and
estimate the sample distribution for the merged set using the method
of ranks, as shown in figure (\ref{powerlaw}).
We fit the resulting distribution to the functional form
\begin{equation}
    P(\delta) = \frac{A}{(x_0 + \delta)^\beta},
    \label{probDist}
\end{equation}
$A$ is set by the normalization, and is a simple function of $x_0$
and $\beta$.  Fitting this to the entire sample (both buys and
sells) gives $x_0 = 7.01 \pm 0.05$, and $\beta = 1.491 \pm 0.001$.
Buys and sells gave similar values for the exponent, i.e. $\beta =
1.49$ in both cases.  Since these error bars based on goodness of
fit are certainly overly optimistic, we also tested the stability
of the results by fitting buys and sells separately on the first
and last half of the sample, which gave values in the range $1.47
< \beta < 1.52$.  Furthermore, we checked whether there are
significant differences in the estimated parameters for stocks
with high vs. low order arrival rates.  The results ranged from
$\beta=1.5$ for high to $\beta=1.7$ for low arrival rates, but for
the low arrival rate group we do not have high confidence in the
estimate.

As one can see from the figure, the fit is reasonably good. The
power law is a good approximation across more than two decades,
for relative limit prices ranging from about $10-2000$ ticks.  For
British stocks ticks are measured either in pence, half pence, or
quarter pence; in the former case, 2000 ticks corresponds to about
twenty pounds. Given the low probability of execution for orders
with such high relative limit prices this is quite surprising.
(For Vodafone, for example, the highest relative limit price that
eventually resulted in a transaction was 240 ticks). The value of
the exponent $\beta \approx 1.5$ implies that the mean of the
distribution exists, but its variance is formally infinite.  Note
that because normalized power law distributions are scale free,
the asymptotic behavior does not depend on units, e.g. ticks vs.
pounds.  There appears to be a break in the power law at about
2000 ticks, with sell orders deviating above and buy orders
deviating below.  A break at roughly this point is expected for
buy orders due to the fact that $p=0$ places a lower bound on the
limit price.  For a stock trading at 10 pounds, for example, with
a ticksize of a half pence, 2000 ticks is the lowest possible
relative limit price for a buy order.  The reason for a
corresponding break for sell orders is not so obvious, but in view
of the extreme low probability of execution, is not surprising.
It should also be kept in mind that the number of events in the
extreme tail is very low, so this could also be a statistical
fluctuation.

\begin{figure}[ptb]
   \begin{center}
   \includegraphics[scale=0.37,angle=-90]{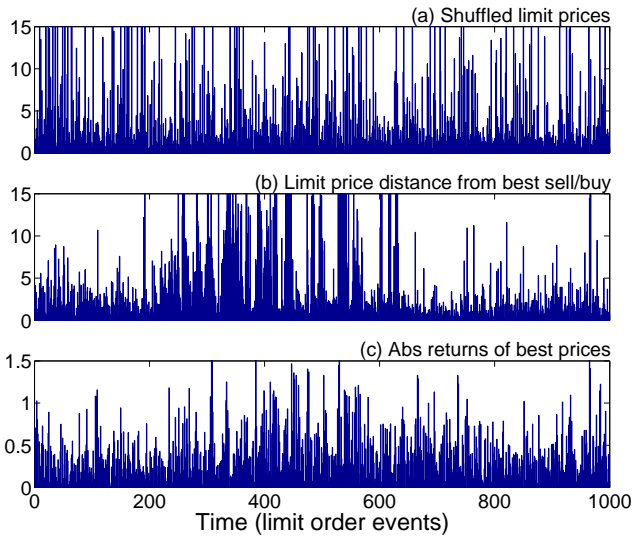}
   \caption{
(a) Time series of randomly shuffled values of $\delta(t)$ for
stock Barclays Bank.  (b) True time series $\delta(t)$. (c) The
absolute value of the change in the best price between each event
in the $\delta(t)$ series. } \label{timeseries}
\end{center}
\end{figure}

The time series of relative limit prices also has interesting temporal
structure.  This is apparent to the eye, as seen in figure
(\ref{timeseries}b), which shows the average relative limit price
$\bar{\delta}$ in intervals of approximately 60 events for Barclays
Bank.  For reference, in figure (\ref{timeseries}a) we show the same
series with the order of the events randomized.  Comparing the two
suggests that the large and small events are more clustered in the
real series than in the shuffled series.

\begin{figure}[ptb]
   \begin{center}
   \includegraphics[scale=0.37, angle=-90]{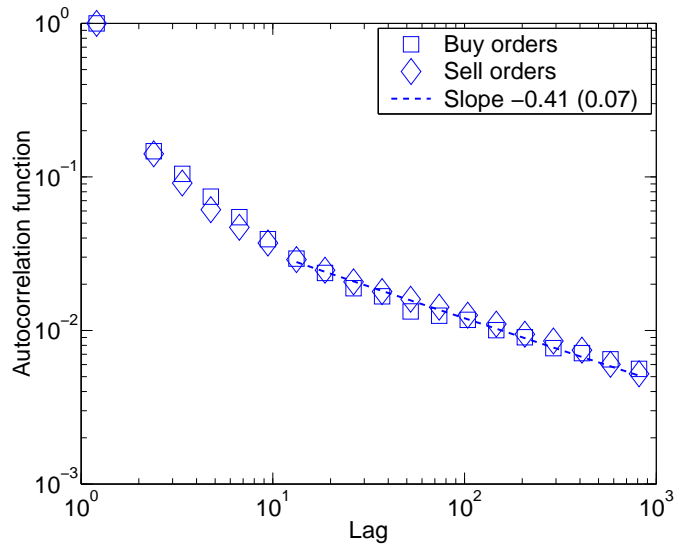}
   \caption{The autocorrelation of the
   time series of relative limit prices $\delta$, averaged across all
   50 stocks in the sample, and smoothed across different lags.  This
   is computed in tick time, i.e., the x-axis indicates the number of
   events, rather than a fixed time.}
  \label{autocor}
  \end{center}
\end{figure}

This temporal structure appears to be described by a slowly decaying
autocorrelation function, as shown in figure (\ref{autocor}).
Since the second moment of the unconditional distribution does not
exist there are potential problems in computing the autocorrelation
function.  The standard deviations in the denominator formally do not
exist, and the terms in the numerator can be slow to converge.  To
cope with this we have imposed a cutoff at 1000 ticks, averaged across
all 50 stocks in our sample, and smoothed the autocorrelation function
for large lags (where the statistical significance drops).  The
resulting average autocorrelation function decays asymptotically as a
power law of the form $C(\tau) \sim \tau^{-\gamma}$, with $\gamma
\approx 0.4$, indicating that relative limit price placement is quite
persistent with no characteristic time scale.  In the figure we have
computed the autocorrelation function in tick time, i.e. the lags
correspond to the event order.  This means that low volume stocks have
longer real time intervals than high volume stocks.  We have also
obtained a similar result using real time, by computing the mean limit
price $\bar{\delta}$ in thirteen minute intervals (merging daily
boundaries).  In this case the behavior is not quite as regular but is
still qualitatively similar.  We still see a slowly decaying power law
tail, though with a somewhat lower exponent (roughly 0.3).  The
autocorrelations are quite significant even for lags of 1000,
corresponding to about 8 days.  Roughly the same behavior is seen for
buy and sell orders, and for the first ten months and the last ten
months of the sample.  We computed error bars for this result by
randomly shuffling the timeseries 100 times, and computing the 2.5 and
97.5 percent quantiles of the sample autocorrelation for each lag.
This gives error bars at roughly $\pm 10^{-3}$.

One consequence of such a slowly decaying autocorrelation is the slow
convergence of sample distributions to their limiting distribution.
If we generate artificial IID data with equation (\ref{probDist}) as
its unconditional distribution, the sample distributions converge very
quickly with only a few thousand points.  In contrast for the real
data, even for a stock with 200,000 points the sample distributions
display large fluctuations.  When we examine subsamples of the
real data, the correlations in the deviations across subsamples are
obvious and persist for long periods in time, even when there is no
overlap in the subsamples.  We believe that the slow convergence of
the sample distributions is mainly due to the long range temporal
dependence in the data.

To get some insight into the possible cause of the temporal
correlations, we compare the time series of relative limit prices to
the corresponding price volatility.  The price volatility is measured
as $v(t) = \left|
\log ( b(t) / b(t-1)) \right|$, where $b(t)$ is the best bid for buy orders
or the best ask for sell limit orders.  We show a typical volatility
series in figure (\ref{timeseries}c).  One can see by eye that epochs
of high limit price tend to coincide with epochs of high volatility.

\begin{figure}[ptb]
   \begin{center}
   \includegraphics[scale=0.37, angle=-90]{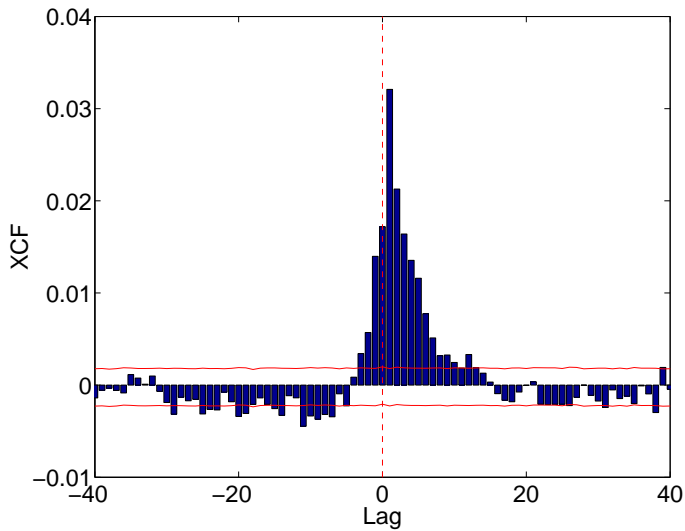}
   \caption{The cross autocorrelation of the time series of relative
   limit prices $\delta(t)$ and volatilities $v(t-\tau)$, averaged across
   all 50 stocks in the sample.}
    \label{crossautocor}
    \end{center}
\end{figure}

To help understand the possible relation between volatility and
relative limit price we calculate their cross-autocorrelation.
This is defined as
\begin{equation}
XCF(\tau) = \frac{\langle v(t - \tau) \delta(t) \rangle - \langle v(t)
\rangle \langle \delta(t) \rangle}{{\sigma_v}
{\sigma_\delta}},
\end{equation}
where $\langle \cdot \rangle$ denotes a sample average, and $\sigma$
denotes the standard deviation.  We first create a series of the average
relative limit price and average volatility over 10 minute intervals.
We then compute the cross-autocorrelation function and average over
all stocks.  The result is shown in figure (\ref{crossautocor}).

We test the statistical significance of this result by testing against
the null hypothesis that the volatility and relative limit price are
uncorrelated.  To do this we have to cope with the problem that the
individual series are highly autocorrelated, as demonstrated in figure
(\ref{autocor}), and the 50 series for each stock also tend to be
correlated to each other. To solve these problems, we construct
samples of the null hypothesis using a technique introduced in
reference \cite{Theiler92}.  We compute the discrete Fourier transform
of the relative limit price time series.  We then randomly permute the
phases of the series, and perform the inverse Fourier transform.  This
creates a realization of the null hypothesis, drawn from a
distribution with the same unconditional distribution and the same
autocorrelation function.  Because we use the same random permutation
of phases for each of the 50 series, we also preserve their
correlation to each other.  We then compute the cross autocorrelation
function between each of the 50 surrogate limit price series and its
corresponding true volatility series, and then average the results.
We then repeat this experiment 300 times, which gives us a
distribution of realizations of averaged sample cross-autocorrelation
functions under the null hypothesis. This procedure is more
appropriate in this case than the standard moving block bootstrap,
which requires choice of a timescale and will not work for a series
such as this that does not have a characteristic timescale. The 2.5\%
and 97.5\% quantile error bars at each lag are denoted by the two
solid lines near zero in figure (\ref{crossautocor}).

From this figure it is clear that there is indeed a strong
contemporaneous correlation between volatility and relative limit
price, and that the result is highly significant.  Furthermore, there
is some asymmetry in the cross-autocorrelation function; the peak
occurs at a lag of one rather than zero, and there is more mass on the
right than on the left.  This suggests that there is some tendency for
volatility to lead the relative limit price.  This implies one of
three things: (1) Volatility and limit price have a common cause, but
this cause is for some reason felt later for the relative limit price;
(2) the agents placing orders key off of volatility and correctly
anticipate it; or, more plausibly, (3) volatility at least partially
causes the relative limit price.  Note that this suggests an
interesting feedback loop: Holding other aspects of the order
placement process constant, an increase in the average relative limit
price will lower the depth in the limit order book at any particular
price level, and therefore increase volatility.  Since such a feedback
loop is unstable, there are presumably nonlinear feedbacks of the
opposite sign that eventually damp it.  Nonetheless, such a feedback
loop may potentially contribute to creating clustered volatility.

One of the most surprising aspects of the power law behavior of
relative limit price is that traders place their orders so far away
from the current price.  As is evident in figure (\ref{powerlaw}),
orders occur with relative limit prices as large as 10,000 ticks (or
25 pounds for a stock with ticks in quarter pence).  While we have
taken some precautions to screen for errors, such as plotting the data
and looking for unreasonable events, despite our best efforts, it is
likely that there are still data errors remaining in this
series. There appears to be a break in the merged unconditional
distribution at about 2000 ticks; if this is statistically
significant, it suggests that the very largest events may follow a
different distribution than the rest of the sample, and might be
dominated by data errors.  Nonetheless, since we know that most of the
smaller events are real, and since we see no break in the behavior
until roughly $\delta \approx 2000$, errors are highly unlikely to be
the cause of the power law behavior seen for $\delta < 2000$.

The conundrum of very large limit orders is compounded by consideration
of the average waiting time for execution as a function of relative
limit price.  We intend to investigate the dependence of the waiting
time on the limit price in the future, but since this requires
tracking each limit order, the data analysis is more difficult.  We
have checked this for one stock, Vodafone, in which the largest
relative limit price that resulted in an eventual trade was $\delta =
240$ ticks. Assuming other stocks behave similarly, this suggests that
either traders are strongly over-optimistic about the probability of
execution, or that the orders with large relative limit prices are
placed for other reasons.

Since obtaining our results we have seen a recent preprint by Bouchaud
et al. \cite{Bouchaud02} analyzing three stocks on the Paris Bourse
over a period of a month.  They also obtain a power law for
$P(\delta)$, but they observe an exponent $\beta \approx 0.6$, in
contrast to our value $\beta \approx 1.5$.  We do not understand why
there should be such a discrepancy in results.  While they analyze
only three stock-months of data, whereas we have analyzed roughly 1050
stock-months, their order arrival rates are roughly 20 times higher
than ours, and their sample distributions appear to follow the power
law scaling fairly well.

One possible explanation is the long-range correlation.  Assuming the
Paris data show the same behavior we have observed, the decay in the
autocorrelation is so slow that there may not be good convergence in a
month, even with a large number of samples.  The sample exponent
$\hat{\beta}$ based on one month samples may vary with time, even if
the sample distributions appear to be well-converged.  It is of course
also possible that the French behave differently than the British, and
that for some reason the French prefer to place orders much further
from the midpoint.

Our original motivation for this work was to model price formation
in the limit order book, as part of the research program for
understanding the volatility and liquidity of markets outlined in
reference \cite{Daniels01}.  $P(\delta)$ is important for price
formation, since where limit orders are placed affects the depth
of the limit order book and hence the diffusion rate of prices.
The power law behavior observed here has important consequences
for volatility and liquidity that will be described in a future
paper.

Our results here are interesting for their own sake in terms of human
psychology.  They show how a striking regularity can emerge when human
beings are confronted with a complicated decision problem.  Why should
the distribution of relative limit prices be be a power law, and why
should it decay with this particular exponent?  Our results suggest
that the volatility leads the relative limit price, indicating that
traders probably use volatility as a signal when placing orders.  This
supports the obvious hypothesis that traders are reasonably aware of
the volatility distribution when placing orders, an effect that may
contribute to the phenomenon of clustered volatility.  Plerou et
al. have observed a power law for the unconditional distribution of
price fluctuations \cite{Plerou99}.  It seems that the power law for
price fluctuations should be related to that of relative limit prices,
but the precise nature and the cause of this relationship is not
clear.  The exponent for price fluctuations of individual companies
reported by Plerou et al. is roughly $3$, but the exponent we have
measured here is roughly $1.5$.  Why these particular exponents?
Makoto Nirei has suggested that if traders have power law utility
functions, under the assumption that they optimize this utility, it is
possible to derive an expression for $\beta$ in terms of the exponent
of price fluctuations and the coefficient of risk aversion.  However,
this explanation is not fully satisfying, and more work is needed
\cite{Nirei02}.  At this point the underlying cause of the power law
behavior of relative limit prices remains a mystery.

\begin{acknowledgments}
We would like to thank Makoto Nirei, Paolo Patelli, Eric Smith, and
Spyros Skouras for valuable conversations and Marcus Daniels for
valuable technical support. We also thank the McKinsey Corporation,
Credit Suisse First Boston, Bob Maxfield, and Bill Miller for
supporting this research.

\end{acknowledgments}


\begin{thebibliography}{9}
\bibitem{foot1}
We have made a somewhat arbitrary choice in defining the ``best
price''.  An obvious alternative would have been to choose the
best ask as the reference price for buy orders, and the best bid
as the reference price for sell orders.  This would have the
advantage that it would have automatically included orders placed
inside the interval between the bid and ask (the spread), which
are discarded in the present analysis.  We chose the reference
prices as we did under the assumption that the best buy price is
the natural reference for a buy order, and similarly for sell
orders.  The choice of reference price does not seem to make a
large difference in the tail; for large $\delta$ it leads to
results that are essentially the same.
\bibitem{foot3}
For orders placed inside the spread we also seem to observe power
law behavior, but with less statistical significance.
\bibitem{foot4}
To avoid overcrowding, we have averaged together nearby bins,
which is why the plots appear to violate the normalization
condition.
\end{thebibliography}
\end{document}